\documentclass[12pt]{emulateapj}

\usepackage[utf8]{inputenc}
\usepackage[usenames,dvipsnames]{color}
\usepackage{natbib}
\usepackage{graphicx}

\def\d{{\rm d}}

\newcommand{\beq}{\begin{equation}}
\newcommand{\eeq}{\end{equation}}

\def\mgii{\ion{Mg}{2}}
\def\Mgii{\ion{Mg}{2}}

\begin{document}

\title{Clustering-based redshift estimation:\\
method and application to data}

\shorttitle{Clustering-based redshifts}
\shortauthors{M{\'e}nard et al.}

\author{
Brice M{\'e}nard\altaffilmark{1,2,3}, 
Ryan Scranton\altaffilmark{4},
Samuel Schmidt\altaffilmark{4},\\
Chris Morrison\altaffilmark{4},
Donghui Jeong\altaffilmark{1},
Tamas Budavari\altaffilmark{1},
Mubdi Rahman\altaffilmark{1}
}
\altaffiltext{1}{Department of Physics \& Astronomy, Johns Hopkins University, 3400 N. Charles Street, Baltimore, MD 21218, USA}
\altaffiltext{2}{Institute for the Physics and Mathematics of the Universe, Tokyo University, Kashiwa 277-8583, Japan}
\altaffiltext{3}{Alfred P. Sloan Fellow}
\altaffiltext{4}{Department of Physics, University of California, One Shields Avenue, Davis, CA 95616, USA}

\submitted{Submitted to MNRAS}

\begin{abstract}
We present a data-driven method to infer the redshift distribution of an arbitrary dataset based on spatial cross-correlation with a reference population and we apply it to various datasets across the electromagnetic spectrum to show its potential and limitations. Our approach advocates the use of clustering measurements on all available scales, in contrast to previous works focusing only on linear scales. We also show how its accuracy can be enhanced by optimally sampling a dataset within its photometric space rather than applying the estimator globally. We show that the ultimate goal of this technique is to characterize the mapping between the space of photometric observables and redshift space as this characterization then allows us to infer the clustering-redshift p.d.f. of a single galaxy. We apply this technique to estimate the redshift distributions of luminous red galaxies and emission line galaxies from the SDSS, infrared sources from WISE and radio sources from FIRST. We show that consistent redshift distributions are found using both quasars and absorber systems as reference populations. This technique brings valuable information on the third dimension of astronomical datasets. It is widely applicable to a large range of extra-galactic surveys. 
\end{abstract}

\keywords{redshift -- clustering}

\section{Introduction}

Observations of the sky are inherently a two-dimensional measurement of electromagnetic flux density as a function of angular position. For astrophysical studies, however, one usually needs the knowledge of three-dimensional positions for example to convert an angle into a physical scale or a brightness into a luminosity. This has been a long-standing limitation in astronomy.

On extragalactic scales, distances are usually inferred from redshift measurements using the knowledge of the expansion history of the Universe. A redshift can be directly measured from observations when one can detect and identify a high-contrast spectroscopic feature. Consequently, robust redshift measurements require spectroscopic observations of sources with emission or absorption lines or spectral break, at a sufficient resolution. Such observations are usually expensive and restricted to bright objects; for example, the Sloan Digital Sky Survey \citep[SDSS;][]{abazajain09} has imaged about 100 million galaxies, but only of order 1\% have been followed-up spectroscopically, most of which are bright and nearby. 
For the vast majority of galaxies, distance estimates rely on so-called ``photometric'' redshifts. They use observed broadband colors to probe the overall spectral energy distribution (SED) of a source. Thus, they rely of qualitatively different information. Photometric redshift estimation suffers from a number of limitations: intrinsic degeneracies between colors and redshifts, unrealistic SED templates, dust reddening, etc. Despite such limitations, however, all upcoming imaging surveys rely on photometric redshifts. With deeper surveys of the sky and access to new wavelength ranges from space, the lack of robust distance estimates is becoming a limitation. Moreover, given the rate at which modern surveys are imaging the sky, the fraction of objects for which we have spectra \emph{decreases} with time. Consequently, alternative techniques should be explored to estimate cosmological redshifts.\\

Redshift inference can be done from a different angle where the estimation is not based on source colors but instead makes use of their angular clustering with a reference or a set of reference populations for which redshifts are well determined. Even though such a technique is currently not being applied, the underlying idea has been discussed for several decades and in a few cases applications to data have pointed out some of its potential. Already thirty five years ago, Seldner \& Peebles (1979) attempted to understand the redshift (and therefore the nature) of quasars by measuring their angular cross-correlations with available galaxy samples. Later on, Phillipps \& Shanks (1987) used counts of faint photometric sources around galaxies with well defined redshifts to obtain an estimate of the galaxy luminosity function at fainter magnitudes. A decade later, Landy, Szalay \& Koo (1996) showed that a combination of auto- and cross-correlations between two populations of galaxies can be used to test whether a significant fraction of the objects from one sample do not overlap in redshift with the other one. At the beginning of the Millenium, requirements for planned photometric surveys designed to constrain the properties of dark energy (which are not met by the photometric redshift techniques currently available) provided some motivation to explore the potential of clustering-based redshift inference more thoroughly. Schneider et al. (2006) first presented a formalism aimed at using clustering information to estimate the accuracy with which photometric redshifts can be inferred and in particular characterize the fraction of catastrophic outliers. Attempting to estimate the redshift distribution of the NVSS radio survey, Ho et al. (2008) showed that useful constraints can be obtained using a combination of spatial auto- and cross-correlations with a few spectroscopic samples. Based on a similar set of observables, Newman (2008) and Matthews \& Newman (2010) presented a method to infer the redshift distribution of a large sample of galaxies using an iterative technique. Finally, expanding on this work, McQuinn \& White (2013) showed how to optimize the corresponding statistical estimator and improve the power of such a procedure.

Surprisingly, more than fifteen years after the encouraging results obtained by Landy et al. (1996), more than five years after a series of theoretical papers mentioned above and a number of studies pointing out how such techniques could improve cosmological experiments (e.g. de Putter et al. 2013) this direction of research has stayed at the level of a theoretical idea and has not led to the promised advances in redshift estimation. None of the proposed techniques has become a generic tool used by the community, applications to real datasets have been largely missing and photometric redshifts are still the only avenue to estimate redshifts when spectroscopic data is unavailable.

In this paper we show that this situation can be changed if we approach the problem differently. We present a practical method to efficiently propagate statistical redshift information from a (small) sample of sources with known redshifts to other objects for which we only have angular positions using information extracted from spatial clustering. Taking into account some of the limitations and challenges involved with real data, we present a method designed to be directly applicable to existing datasets. As we are building a new tool from scratch we are not directly aiming at percent-level accuracy (which was the goal of a number of the theoretical papers written on the subject over the past five years). In contrast to previous proposals, we advocate for the use of small-scale clustering measurements (i.e. in the non-linear regime), a sampling done locally in the photometric space as opposed to applying the method to an entire dataset and we avoid using information from auto-correlation functions which are more subject to systematic effects than cross-correlations with real data.
This technique is ultimately aimed at characterizing the mapping between the space of observables accessible from the photometry to redshift space. We also point out that having characterized this mapping the technique \emph{can} provide us with the redshift p.d.f. of individual galaxies, similarly to photometric redshift estimation. Finally we demonstrate the power of this technique by applying it to existing datasets across the electromagnetic spectrum, from the optical to the radio range (where photometric redshifts cannot even be defined) and estimate the corresponding redshift distributions. In a companion paper \citep{schmidt13} we present results from numerical simulations to test the robustness and limits of our redshift inference method when applied to realistic distributions of dark matter halos and galaxies, and in Rahman et al. (2014) we will show how clustering-based redshifts compare to spectroscopic redshifts for galaxies selected from the SDSS.

\section{Clustering-based redshift estimation}

\subsection{The covariance of the sky}

Electromagnetic observations of the sky consist of a measurement of flux density $F_\lambda$ as a function of angular position. We denote the flux density fluctuation at a location $\phi$ and wavelength $\lambda$ as $\delta F_{\lambda}(\phi)$. The generic covariance of the extragalactic sky is given by
\begin{equation}
C_{\rm obs}(\lambda_1,\lambda_2,\theta) = 
\langle \delta F_{\lambda_1}(\phi)\;\delta F_{\lambda_2}(\phi+\theta)\rangle_\phi~.
\end{equation}
This quantity is uniquely defined and provides us with statistical information on the extragalactic sky as a function of position and wavelength. 

If we have access to a population of objects whose spatial distribution characterized by the density contrast $\delta(\vec r)$ is located within a narrow redshift bin centered on $z_0$, we can use it to probe a projection of the observed flux density fluctuation $\delta F_{\lambda}$:
\begin{equation}
C_{\rm obs}(\lambda,r_p,z) = 
\langle \delta(z_0)\;\delta F_{\lambda}(r_p)\rangle\;.
\label{eq:main}
\end{equation}
Here we note that the flux fluctuation $\delta F_\lambda$ is not restricted to discrete objects, like galaxies, quasars or GRBs, but can also be a continuous field such as the infrared background or a millimetric temperature map. Eq.~\ref{eq:main} indicates that the observable $C_{\rm obs}$ can be used to extract some information on the redshift distribution of an arbitrary dataset. A lack of correlation can be used to test for the absence of objects in $\delta F_{\lambda}$ at redshift $z_0$.

We note that the calibration of photometric redshifts with observed spectra makes use of the quantity $C_{\rm obs}$ through correlations between a known redshift and the observable $\delta F_{\lambda}$ (or similarly a color), but restricts the spatial dependence to $r_p=0$. One of the main limitations of photometric redshifts is due to the fact that the correlation $C_{\rm obs}(\lambda,r_p=0,z)$ measured for different objects at different redshifts can lead to the same amplitude which gives rise to degeneracies between redshifts and colors. 

An important point of this paper is that the (projected) environment of a source can be treated as an observable which, in a statistical context, can be a powerful indicator of its properties, including its redshift. Due to the existence of overlapping objects along the line-of-sight, the projected environment is often a noise-dominated quantity. However, if one is interested in estimating the redshift of an ensemble of objects, the mean projected environment can become a signal-dominated quantity and a useful source of information. We now show how to make use of this information to infer the redshift distribution of a population for which we only know the angular positions on the sky.

\subsection{Redshift inference from spatial clustering}
\label{sec:inference}

\subsubsection{Ideal case}

Let us consider two populations of extragalactic objects: (i) a \emph{reference} population for which we know the angular positions and redshifts of each object. This population is characterized by a redshift distribution $\d{\rm N_r}/\d z$ and a mean surface density $n_r$ and a total number of sources $N_r$; and (ii) an \emph{unknown} population for which angular positions are known but redshifts are not. Similarly, this population is characterized by the quantities $\d{\rm N_u}/\d z$, $n_u$ and $N_u$. We first consider an ideal case in which all the unknown sources are at the same redshift $z_0$:
\begin{equation}
\frac{\rm \d N_u}{\d z} = {\rm N_u}\,\delta_D(z-z_0)\;.
\end{equation}
As shown below, this is a regime where clustering-based redshift estimation provides us with an unbiased and accurate estimate of ${\rm \d N_u /\d}z$. The next section will show the interest of exploring the neighborhood of this regime. Even if our clustering-based estimator is no longer unbiased, there might exist a regime in which the final accuracy is sufficient for many astrophysical purposes.

To probe the redshift distribution of the unknown sample we split the reference population in redshift bins $\delta z_i$ and for each subsample $i$ we measure its angular or spatial correlations with the unknown population $w_{ur}(\theta,z_i)$:
\begin{equation}
w_{ur}(\theta,z_i) = \frac{ \langle n_u(\theta,z_i) \rangle_r }{n_{u}} - 1\,,
\label{eq:estimator}
\end{equation}
where $\langle n_u(\theta,z_i) \rangle_r$ denotes the mean density estimate of the unknown sample around reference objects at redshift $z_i$. Given the assumption that all unknown sources are located at $z_0$, we are in a regime where we are only looking for the presence or absence of correlated unknown objects within a reference redshift bin $\delta z_i$. In this case we simply have 
\begin{equation}
{\rm d N_u/\d} z\propto w_{ur}(z_i)\,.
\end{equation}
Once a cross-correlation signal is found the amplitude of the redshift distribution is simply obtained through the normalization
\begin{equation}
\int \d z\,\frac{\rm d N_u}{\d z} = {\rm N_u}\,.
\label{eq:normalization}
\end{equation}
This relation is satisfied if all the objects of the unknown sample are extragalactic \emph{and} if the redshift distribution of the reference population is wide enough to cover the redshift range of the unknown objects. This implies that, in the case of a narrow redshift distribution, it is possible to fully characterize it using clustering information.

At this stage we investigate how to optimize the sensitivity of such an estimator. It is important to note that so far we have not specified how to measure the angular correlation between the unknown population and the set of reference subsamples. Indeed, if $\d{\rm N_u}/\d z\rightarrow {\rm N_u}\,\delta_D(z-z_0)$ we are simply addressing a yes-or-no question whose answer is only limited by the shot noise induced by the finite size of the samples and in some cases cosmic variance. The clustering signal can therefore be measured on any scale and its sensitivity can be maximized by including clustering information from \emph{all} scales available to the measurements. Approaching the problem from this angle does not restrict the analysis to large-scale clustering signals where the galaxy over density field behaves linearly with respect to that of the dark matter, as advocated by previous studies. As a measure of clustering we will consider the integrated cross-correlation function
\begin{equation}
\bar w_{ur}(z) = \int_{\theta_{\rm min}}^{\theta_{\rm max}} \d\theta\, W(\theta)\, w_{ur}(\theta,z)
\label{eq:w_int}
\end{equation}
where $W(\theta)$ is a weight function, whose integral is normalized to unity, aimed at optimizing the overall S/N. As the matter correlation function can often be approximated by a power law over a broad range of scale with $\gamma$ of order unity, we can simply use $W(\theta)\propto \theta^{-\gamma}$. We note that for $\gamma=1$ there is an equal amount of clustering information per logarithmic scale. In order to probe the same range of physical scales as a function of redshift we set $(\theta_{min},\theta_{max})$ to match a fixed range of projected radii $(r_{p,min},r_{p,max})$. We note that as the angular scale becomes comparable to the mean separation between reference objects, number count estimates become correlated and the amount of useful clustering information decreases. In addition, such large-scale estimates are often more subject to systematic effects due to fluctuations in the zero point of the photometry, uncertainties due to Galactic dust extinction effects, etc. Therefore, in practice, we will limit our clustering measurements to scales smaller than several Mpc, which typically correspond to several degrees on the sky. This dramatically contrasts with previous studies using \emph{only} clustering measurements on scales greater than several Mpc. Finally, we set the angular scale $\theta_{min}$ to be always greater than the maximum between the typical size of the sources involved and the point spread function of the corresponding survey. In practice, this typically allows us to measure clustering over more than two orders of magnitude in scale.

It is now interesting to characterize the size of the samples required to use this technique and obtain detectable signals. To do so in a simple manner, we will assume that matter clusters with some scale $r_c$ or $\delta z_c$ in redshift space and not beyond. In this case the signal-to-noise ratio of the measurement of a spatial correlation between a reference subsample selected in the redshift bin $\delta z_i$ and the unknown sample is given by
\begin{eqnarray}
\frac{S}{N}
&\simeq&
\frac{\delta z_c}{\delta z_i}\,
\frac{\bar w_{ur}}{\theta_{max}\sqrt{\pi}}\,
\sqrt{N_{r,i}\,n_u}\nonumber\\
&\simeq&
\frac{\delta z_c}{\sqrt{\delta z_i}}\,
\theta_{max}\,
\sqrt{
\frac{\rm \d N_r}{\d z}
\,n_u}
\label{eq:sn}
\end{eqnarray}
where $N_{r,i}$ is the number of reference objects in the redshift bin $\delta z_i$ and where we have assumed that $\theta_{min}\ll\theta_{max}$. This expression shows that the best strategy is to use reference redshift bins with a size matching that of the correlation length of matter clustering. This maximizes the contrast between the angular correlation measured at the location of the unknown sources and elsewhere. To put this estimate in perspective, we consider parameters representative of the galaxy spectroscopic sample available with the SDSS. Taking the fiducial parameters $\delta z_c=\delta z_i=10^{-3}$, $\theta_{max}=1\,{\rm deg}$, ${\rm \d N_r/\d z}=10^6$, we obtain $S/N\sim 30\,\sqrt{n_u}$, with $n_u$ in units of number of galaxies per square degree. For reference, the number density of photometric galaxies in the survey, selected with $r<21$ is about 3600/deg$^2$ \citep{2000AJ....120.1579Y}. This shows that the clustering redshift technique can be applied to \emph{many} (of order one thousand) subsets of the SDSS photometric sample, provided we can select them so they are located in narrow redshift bins. A narrow beam survey like COSMOS \citep{2007ApJS..172....1S} is similarly appropriate: with a photometric number density of $10^6/{\rm deg^2}$ and about $10^4$ spectroscopic redshifts available, the statistical power of the estimator is high enough to be able to detect a cross-correlation signal for a very large number of subsamples narrowly distributed in redshift space. As shown in Eq.~\ref{eq:sn}, the statistical power depends on the number of pairs between the reference and unknown samples as expected with clustering measurements.

In the general case, Eq.~\ref{eq:estimator} provides us with a robust estimator to precisely locate the redshift range over which an arbitrary population is distributed. It provides us with a \emph{data-driven} approach to test for the presence or absence of sources at a given redshift $z$ and it can be applied to any continuous or discrete dataset. When probing sources for which spectral energy distribution templates are not available (for example because the physics of the objects is not understood) or for which no spectroscopic data is available, the proposed cluster-based redshift estimation provides us with a robust way to infer the presence/absence of sources as a function of redshift, without any assumption.

\subsubsection{Departure from the ideal case}

\begin{figure}[t]
\includegraphics[width=0.49\textwidth]{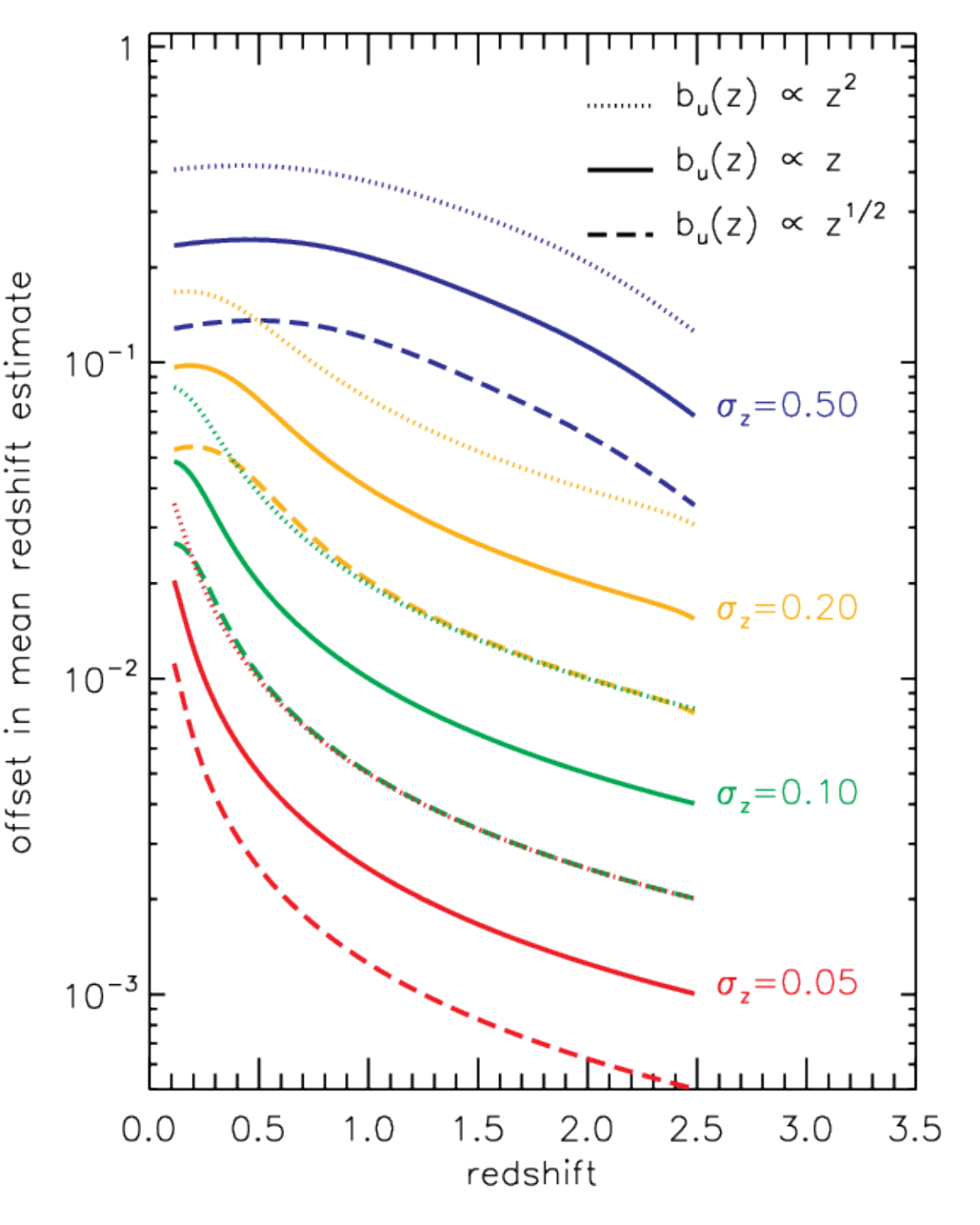}
	\caption{Offset in the estimation of the mean redshift of a sample due to the lack of knowledge of its clustering amplitude ${\overline b_u}(z)$. The figure shows different scenarios: ${\overline b_u}\propto z^{1/2}$, $z$ and $z^2$, for different fiducial populations with redshift distributions characterized by Gaussians with mean redshift $z_0$ and half width $\sigma_z$. For a broad range of parameters considered this shows that the error induced by assuming a non-evolving ${\overline b_u}$ is small enough to allow a large range of astrophysical studies.}
	\label{fig:z_offset}
\vspace{.2cm}
\end{figure}

In the more general case where the unknown population is not located at a single redshift but spread over an interval $\Delta z$, the spatial cross-correlations with the set of reference samples will depend on a number of quantities: the type of unknown and reference objects, their relative clustering amplitude with respect to the dark matter density field, the redshift dependence of the corresponding quantities and the scale over which correlations are considered. By selecting narrow redshift bins $\delta z_i$ of reference objects, the amplitude of the measured angular cross-correlation with the unknown population follows
\begin{equation}
\bar w_{ur}(z_i) \propto
\frac{\rm d N_u}{\d z}(z_i)\,
{\overline b_u}(z_i)\, {\overline b_r}(z_i)\,
\bar w_{\rm DM}(z_i)\,,
\label{eq:w_full}
\end{equation}
where the bar indicates that the quantities have been integrated over a range of scales, according to Eq.~\ref{eq:w_int}. Now we are no longer using the angular correlation to answer a yes-or-no question but we are aiming at constraining the shape of the redshift distribution $\d{\rm N_u}/\d z$. Here it is interesting to comment on several aspects of the above relation:
\begin{itemize}
\item the degree of variation of each term in equation~\ref{eq:w_full} is in general expected to differ. If over the redshift range $\Delta z$ the relative variation of $\d{\rm N_u}/\d z$ dominates over that of ${\overline b_u}(z)$, or in other words if
\begin{equation}
\frac{\d \log {\rm \d N_u/\d}z}{\d z}
\gg
\frac{\d \log {\overline b_u}}{\d z}
\label{eq:recovery_criteria}
\end{equation}
we then approach the regime in which $\d{\rm N_u}/\d z\rightarrow {\rm N_u}\,\delta_D(z-z_0)$ and one can use the method described in the previous section to infer $\d{\rm N_u}/\d z$, but this time only up to some finite accuracy. The amplitude of the expected offset will be described below.
\item We note that in order fully characterize the redshift distribution of the unknown population as advocated in the previous section and normalize its amplitude (Eq.~\ref{eq:normalization}), this only requires the knowledge of the derivates ${\rm \d {\overline b_u}/\d} z$ and ${\rm \d {\overline b_r}/\d} z$. The amplitudes of the two clustering biases are not required.
\item Constraints on the clustering amplitude $\bar b_r$ of the reference sample can in principle be derived from measurements of the autocorrelation function of the reference sample as a function of redshift. 
\begin{equation}
\bar w_{rr}(z) = \bar b_r^2(z)\, \bar w(z)\;.
\end{equation}
While this relation is valid only on scales where galaxies are linearly biased with respect to the dark matter field, the inclusion of smaller scales provides only a modest departure from it. We demonstrate this point in our companion paper (Schmidt et al. 2013) using numerical simulations. 
In addition, we point out that our estimate is based on an average over a wide range of scales which weakens the non-linear effects. Finally, the clustering amplitude of dark matter as a function of redshift is a quantity that is characterized from the theory.
\end{itemize}
The main limitation in estimating $\d{\rm N_u}/\d z$ using Eq~\ref{eq:normalization} \& Eq.~\ref{eq:w_full} originates from the lack of knowledge of the redshift dependence of ${\rm \d {\overline b_u}/\d}z$. Several authors have proposed to constrain this quantity using the measured auto correlation function of the unknown sample, using a redshift averaged value (Ho et al. 2008) or attempting to deproject its redshift dependence through an iterative technique (Newman 2008). Here we propose a different approach. Instead of attempting to characterize this term, we can minimize its contribution to the $\d{\rm N_u}/\d z$ estimator and/or estimate the error induced by approximating the redshift distribution without its contribution. Indeed, it turns out that the error introduced by the lack of information on ${\rm \d {\overline b_u}/\d}z$ is in many cases small enough for this technique to provide useful constraints on redshift distributions. To quantify this effect we consider the following case. Let us assume that, for simplicity, the unknown redshift distribution is represented by a Gaussian distribution $G(z_0,\sigma_z)$ centered on $z_0$ and with a half width $\sigma_z$, i.e. that the redshift distribution of the unknown population roughly extends over a redshift support $\Delta z\sim{\mbox{a few}}\times\sigma_z$. Let us assume that 
\begin{equation}
{\rm {\overline b_u}}(z) \propto z^\alpha\,.
\end{equation}
If we neglect this redshift dependence when using the set of cross-correlation functions to estimate the unknown redshift distribution, i.e. if we simply use
${\rm \d {\overline b_u}/\d}z=0$, the difference between the mean estimated redshift and the true value is given by
\begin{equation}
\langle z\rangle_{\rm est} - z_0 =
\int \d z\, z^{\alpha+1}\,G(z_0,\sigma_z)-\int \d z\, z\,G(z_0,\sigma_z)\,.
\end{equation}
This offset in the inferred mean redshift is shown in Figure~\ref{fig:z_offset} as a function of the mean redshift $z_0$, half width $\sigma_z$ and for three values of $\alpha$. We note that $\alpha\sim1$ is representative of the observed bias evolution for brightness limited samples of low-redshift galaxies 
\citep{2011ApJ...736...59Z}.
%(Zehavi et al., Rahman et al. 2014). 
As expected, the estimated mean redshift will be systematically higher than the real one. Interestingly, we can see the error in the mean redshift is, in many realistic cases, of order several percents, i.e. it can be small enough to allow a large range of astrophysical studies. As an illustration, let us consider a color selected sample of low redshift galaxies. Using a limiting magnitude of $r\sim18$ and a simple color cut $g-i\simeq0.1$, one can select galaxies for which the redshift distribution is relatively well represented by $G(z_0=0.2, \sigma_z=0.05)$. For such a population figure~\ref{fig:z_offset} shows that using a set of angular cross-correlations combined with the overall normalization given in Eq.~\ref{eq:normalization} and using ${\rm \d {\overline b_u}/\d}z=0$ will lead to a mean redshift estimate for which the error is of order one percent. We note that quantities with shallower redshift dependencies, for example $\bar w_{\rm DM}(z)$, will induce biases at levels lower than those illustrated in the figure.

% . . . . . . . . . . . . . . . . . . . . . . . . . . . . . . . . . .
% . . . . . . . . . . . . . . . . . . . . . . . . . . . . . . . . . .

We have shown that, in realistic cases where the redshift distribution of the unknown population is distributed over a range $\Delta z$, combining a set angular cross-correlations with the overall normalization given in Eq.~\ref{eq:normalization} -- and neglecting the bias evolution of the unknown population -- leads to redshift estimations accurate enough for many astrophysical studies.

Finally, we point out that the lack of knowledge of the clustering amplitude is, in general, expected to only affect large-scale modes of the estimated redshift distributions. The method presented above is sensitive to small-scale structure in the redshift distribution of the unknown sample. In other words, for sufficient S/N, we expect it to reveal sudden changes in the redshift distribution of a given sample, for example when massive concentrations of matter are present due to galaxy clusters, walls or filaments aligned in the plane of the sky.

% ------------------------------------------------------------------
\vspace{.2cm}
\subsubsection{Generalization \& Strategy}
\label{sec:strategy}

Redshift estimation based on photometric information can be described as the characterization of the mapping connecting volume elements (or voxels) of the space of photometric observables to redshift space. We note that so-called photometric redshifts characterize this mapping with calibration based on theoretical or observed sets of spectral energy distributions. Our clustering-based estimation aims at characterizing the very same mapping but using spatial correlations.

Typically, the space of photometric observables is characterized by brightness, colors, size, shape and higher-order moments of the light distribution. This space can also include information that is not directly object-based and for example include information on the environment of the sources. The dimensionality of the the space of observable is therefore appreciable. For typical multi-band ground-based surveys it is of order ten. Adding flux measurements over a broad range of wavelength, from the UV to radio, the dimensionality can be increased by another decade. The more parameters are available, the more likely it is to identify regions of the photometric space mapping onto narrow redshift intervals. 

There exists a fundamental mapping between a given space of photometric observables and redshift space. Every photometric voxel $j$ maps onto a redshift distribution of finite extend $\Delta z_j$. Certain regions of this space may map onto multimodal regions of redshift space due to \emph{intrinsic} degeneracies in the mapping itself. There is a limit to how much redshift information can be extracted from the photometry and it is important to realize that it will apply to both photometric redshifts and clustering-based redshifts in the same way. In the case of a photometric voxel mapping onto a multimodal redshift distribution, if selecting subsamples as a function of other photometric dimensions does not break the redshift degeneracy it means that all the available information existing in the mapping between the photometric space and redshift space has been exhausted and photometric information alone does not allow us to differenciate the modes of the distribution.

Given that the accuracy of the proposed redshift inference method becomes higher when considering samples more narrowly distributed in redshift, it implies that the best strategy to use this technique is not to apply it to a sample as a whole (spread over a large redshift range $\Delta z$) but to break it into as many redshift subsamples as possible. Each subsample can be selected by considering a voxel $j$ in the space of observables and apply the proposed technique to obtain an estimate of the redshift support $\Delta z_j$ and/or the modality of the corresponding redshift distribution. If this estimate is too noisy, the cell can be enlarged to increase the number of objects. Having obtained some knowledge of the redshift support $\Delta z_j$ of each voxel allows us to estimate the degree of uncertainty of the redshift distribution inferred for each photometric voxel, as illustrated in Figure~\ref{fig:z_offset}. During this process, voxels with poor mapping onto redshift space, for example due to large $\Delta z_j$ values or multimodal distributions, can be discarded from the final sample of consideration. The final redshift distribution of the parent sample, or a set of voxels with well-defined characterization onto redshift space, is then given by
\begin{equation}
\frac{\rm d N_u}{\d z}
= \sum_{j} \frac{\rm d N_u}{\d z}({\rm voxel\,}j) \,,
\end{equation}
where $j$ is a photometric voxel. 
Our technique advocates for a local sampling of the redshift distribution of an unknown population in the space of its observable parameters in constrast to other methods proposed in the literature, primarily aimed at inferring global distributions.

% ------------------------------------------------------------------
\subsubsection{Reference samples available}
\label{sec:strategy}

% . . . . . . . . . . . . . . . . . . . . . . . . . . . . . . . . . .
\begin{figure}[t]
\vspace{-4cm}
\hspace{-.5cm}
\includegraphics[width=10cm]{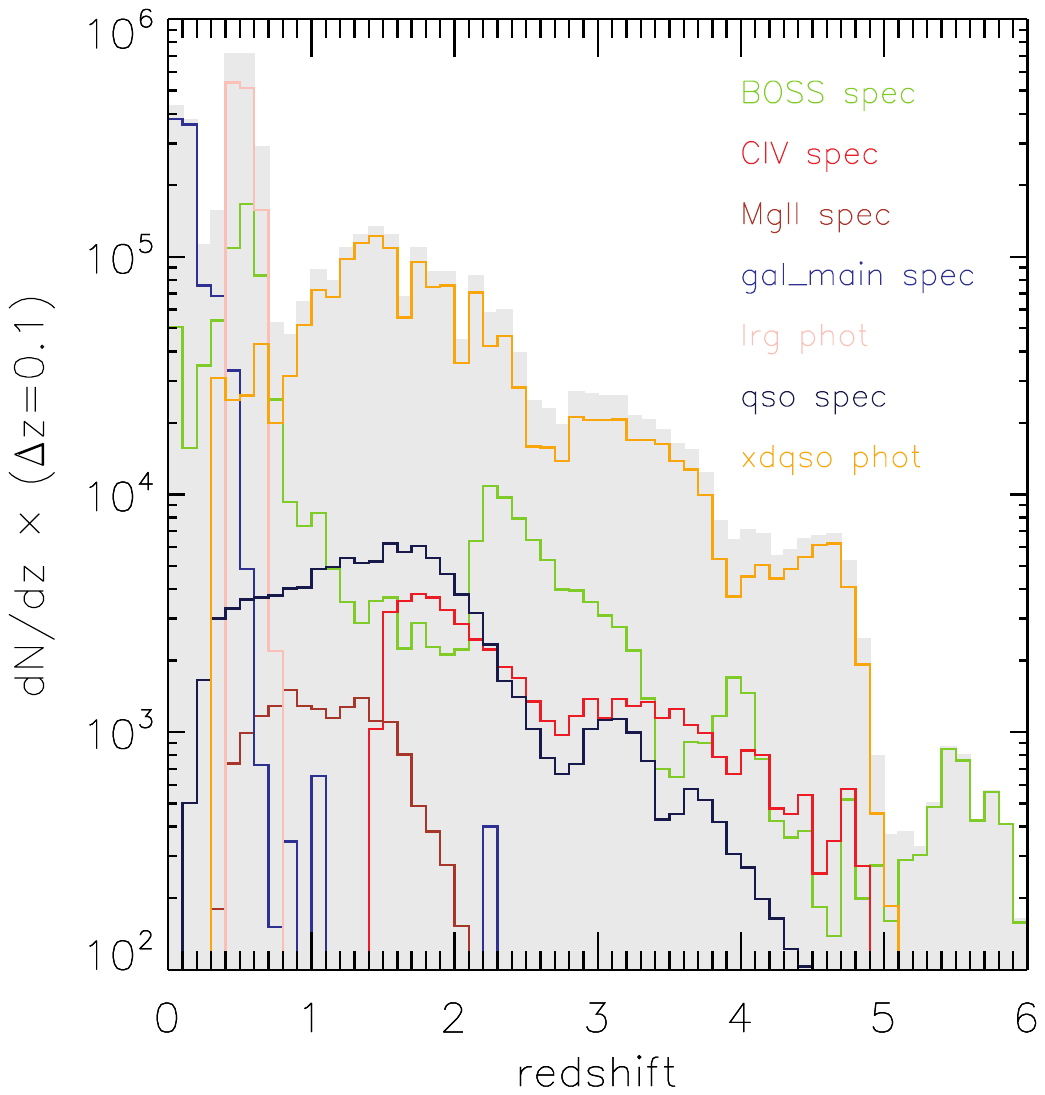} 
	\caption{Compilation of samples from the SDSS for which we have a robust 3d position, either from spectroscopic or photometric redshifts. In this paper we make use of the spectroscopic samples of quasars and \Mgii\ absorbers as shown with the dark blue and brown curves.}
	\label{fig:surveys}
\end{figure}
% . . . . . . . . . . . . . . . . . . . . . . . . . . . . . . . . . .

Interestingly we now have access to a variety of surveys providing us with 3d positions (based on spectroscopic redshifts or, in some cases, sufficiently accurate photometric redshifts), many of which are large enough to be used as reference samples for clustering redshift estimation. As an illustration, we show in figure~\ref{fig:surveys} a compilation of samples drawn from the Sloan Digital Sky Survey \citep[SDSS;][]{abazajain09} for which the redshift distributions are known. The figure includes distributions for galaxies, quasars and absorber systems. As can be seen, the usability criterion given in Eq.~\ref{eq:sn} is met by numerous samples. This figure also shows that different populations can be used to check the consistency of the inferred redshift distributions. In the next section we will make use of the spectroscopic quasar and absorber samples as reference populations. Those are shown with the dark blue and brown curves, respectively. While SDSS quasars are found over the redshift range $0<z<6$, \mgii\ absorbers are only visible in the range $0.4<z<2.2$.

\subsubsection{Gravitational lensing effects}
\label{sec:lensing}

The apparent spatial density of sources in the sky is modulated by gravitational magnification effects due to the matter distribution along the line-of-sight \citep[e.g.,][]{narayan89}. This induces an apparent correlation between populations of objects lying at different redshifts. The amplitude of this effect, also called cosmic magnification, has been estimated by several authors \citep[see][]{bs01} and detected by the large-scale distribution of galaxies by \citet{scranton05} and \citet{menard10}. For sources at high redshift lensed by typical galaxies at $z\sim0.5$, the amplitude of the magnification effect is about 1\% on a scale of one arcminute. In general, this is negligible compared to the signal induced by physical clustering of overlaping samples. In addition, the redshift dependence of the lensing efficiency varies slowly with redshift. The absence of such a signature in the redshift distribution inferred by the spatial cross-correlation technique directly indicates that cosmic magnification effects are not playing a significant role.

\section{Application to data}\label{sec:application}

% . . . . . . . . . . . . . . . . . . . . . . . . . . . . . . . . . .
\begin{figure*}[t]
\vspace{-4cm}
\hspace{-1cm}
\includegraphics[width=10cm]{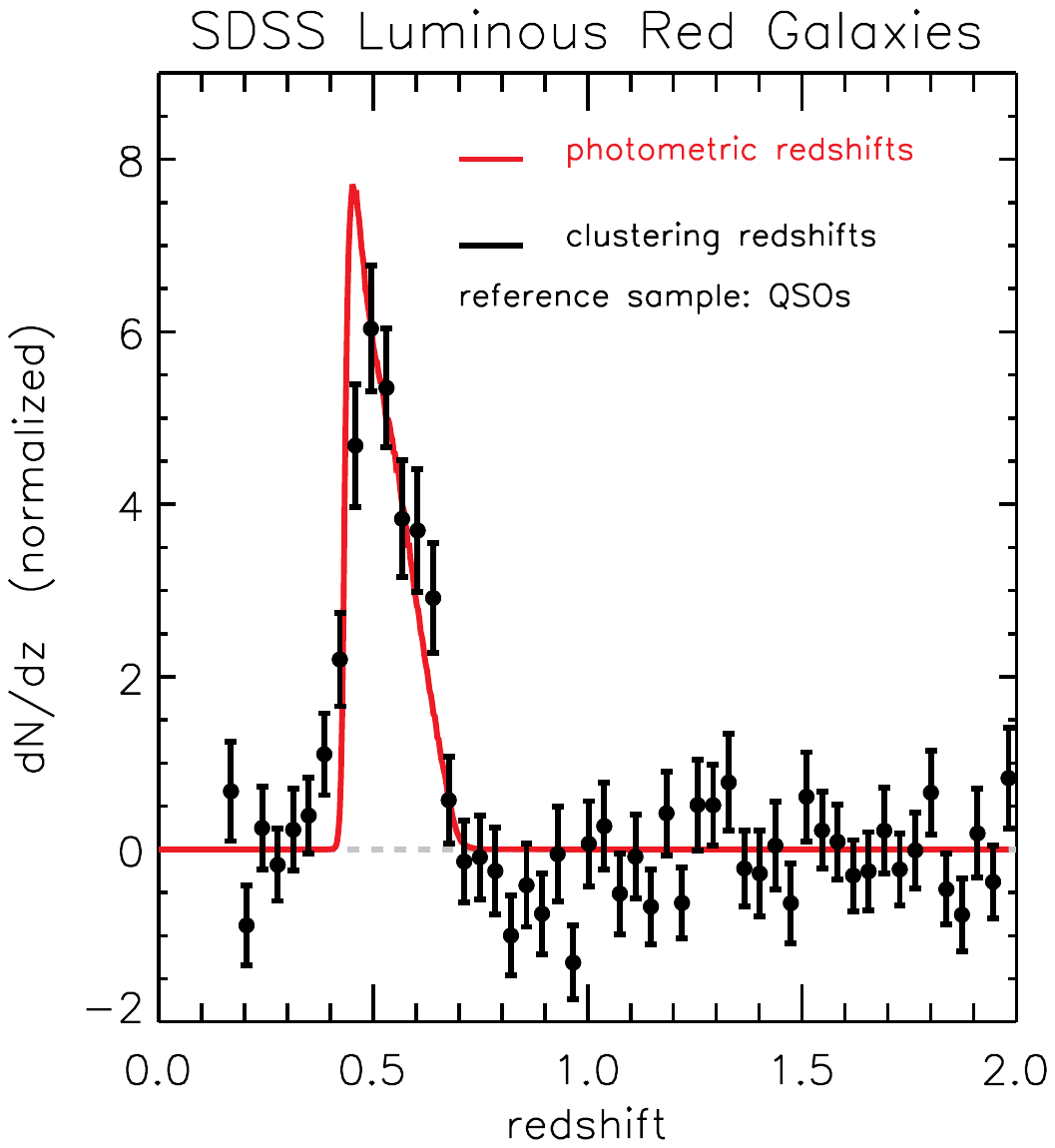}
\includegraphics[width=10cm]{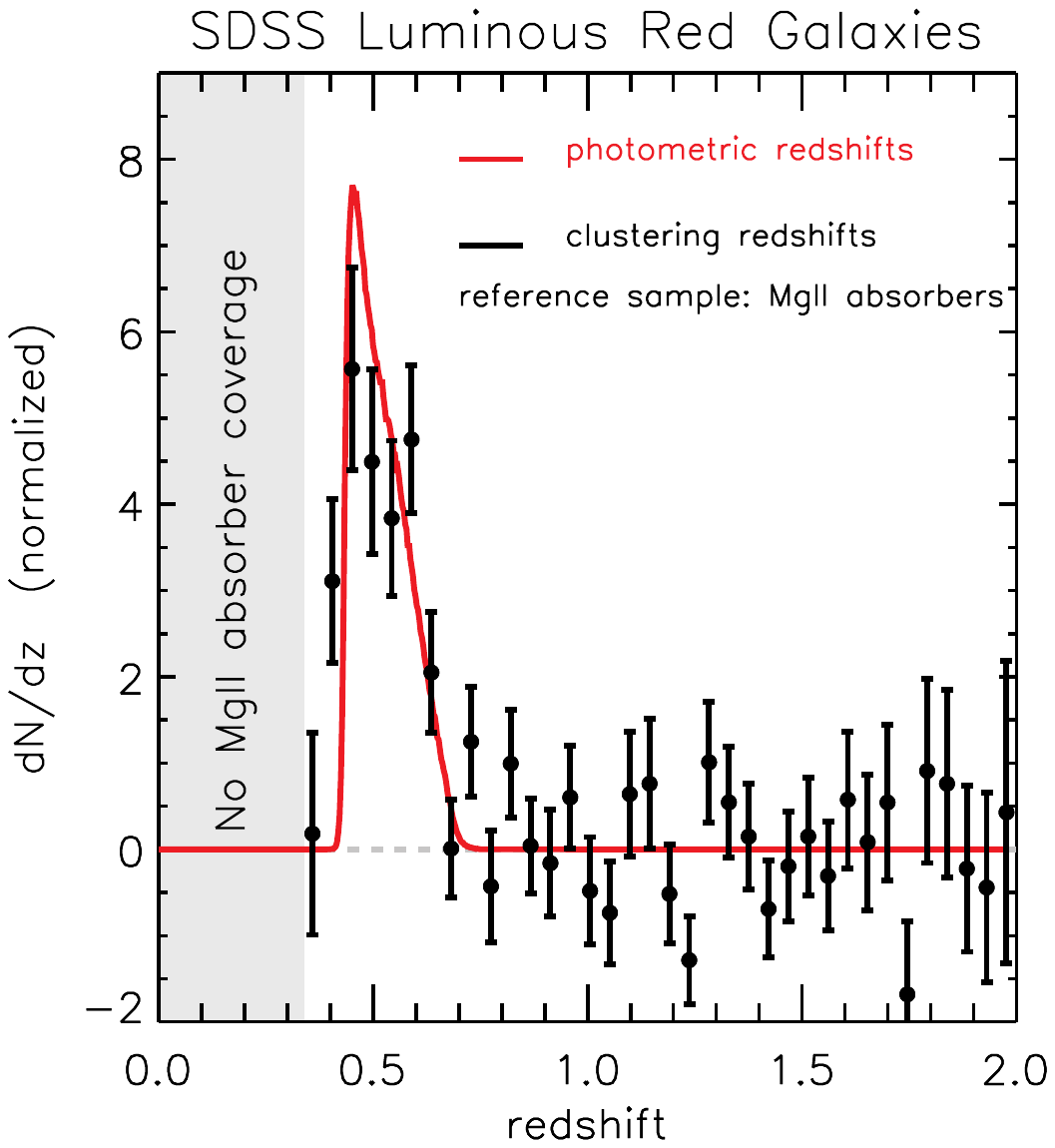}
	\caption{Redshift distributions ${\rm dN/d}z$ (normalized to unity) for Luminous Red Galaxies (LRGs). In both panels the solid red line shows the distribution of LRG photometric redshifts.
    \emph{Left:} cluster-z distribution (black points) obtained by measuring the spatial cross-correlation between LRGs and SDSS quasars.
    \emph{Right:} cluster-z distribution (black points) obtained by measuring the spatial cross-correlation between LRGs and \mgii\ absorbers, spanning the range $0.4<z<2.$}
	\label{fig:LRG}
%\end{figure*}
%%% . . . . . . . . . . . . . . . . . . . . . . . . . . . . . . . . . .

%%% . . . . . . . . . . . . . . . . . . . . . . . . . . . . . . . . . .
%\begin{figure*}[h]
\vspace{-3cm}
\hspace{-1cm}
\includegraphics[width=10cm]{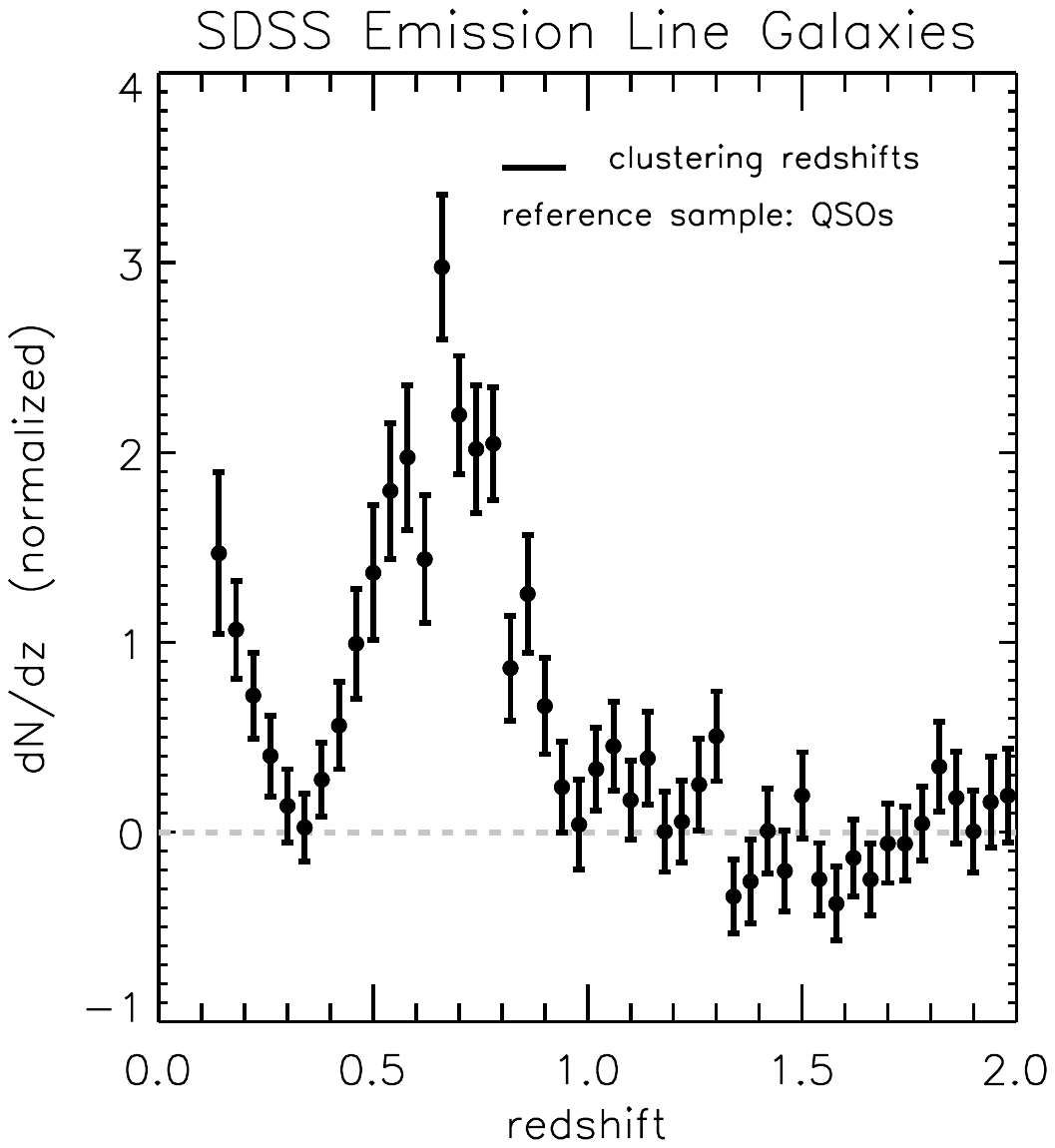}
\includegraphics[width=10cm]{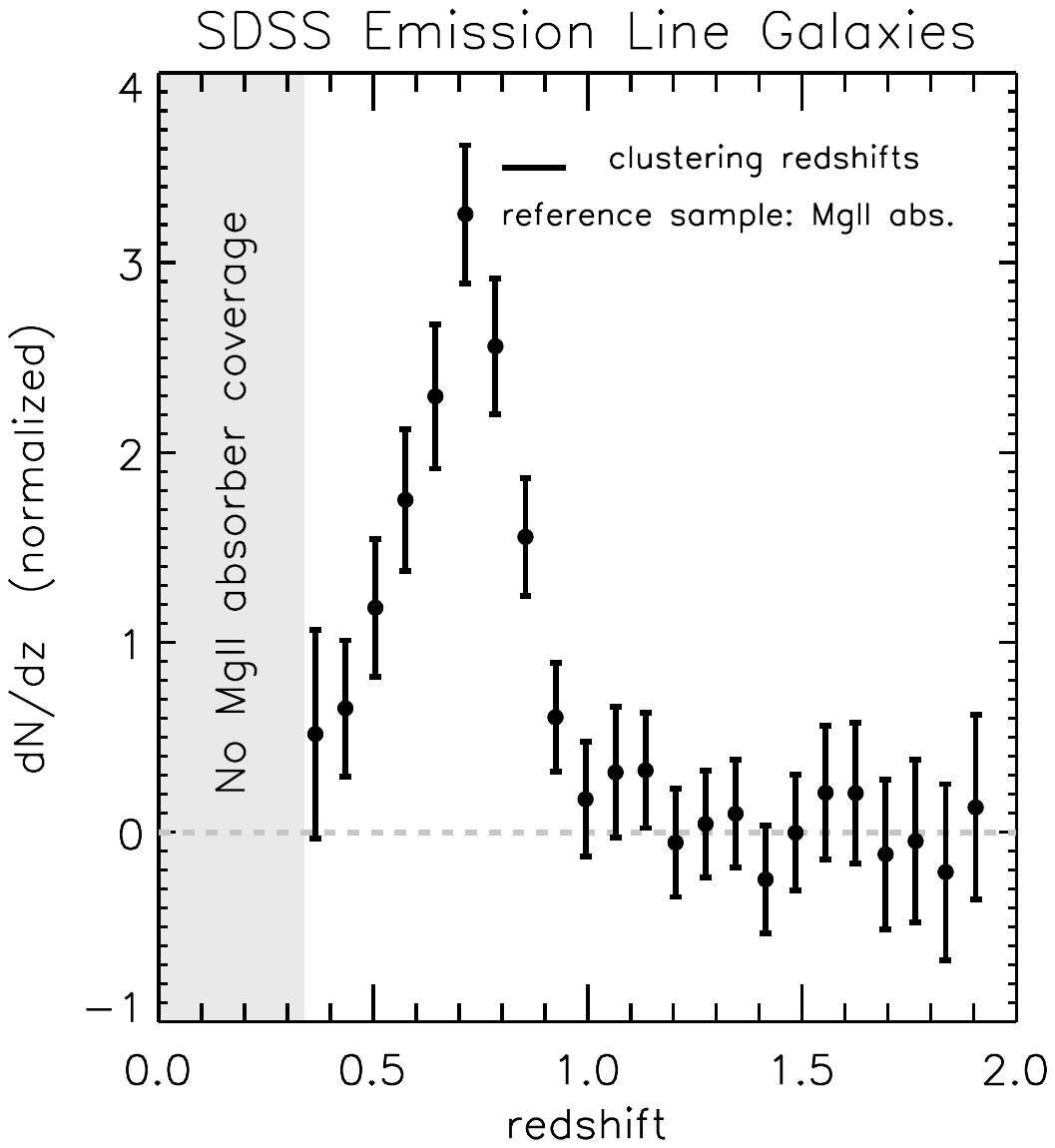}
	\caption{Redshift distributions ${\rm dN/d}z$ (normalized to unity) for Emission Line Galaxies (ELGs) from the SDSS.
        \emph{Left:} cluster-z distribution (black points) obtained by measuring the spatial cross-correlation with SDSS quasars.
    \emph{Right:} cluster-z distribution (black points) obtained by measuring the spatial cross-correlation with \mgii\ absorbers, spanning the range $0.4<z<2.$}
	\label{fig:ELG}
\end{figure*}
% . . . . . . . . . . . . . . . . . . . . . . . . . . . . . . . . . .

% . . . . . . . . . . . . . . . . . . . . . . . . . . . . . . . . . .
\begin{figure*}[t]
\vspace{-3.5cm}
\hspace{-1cm}
\includegraphics[width=10cm]{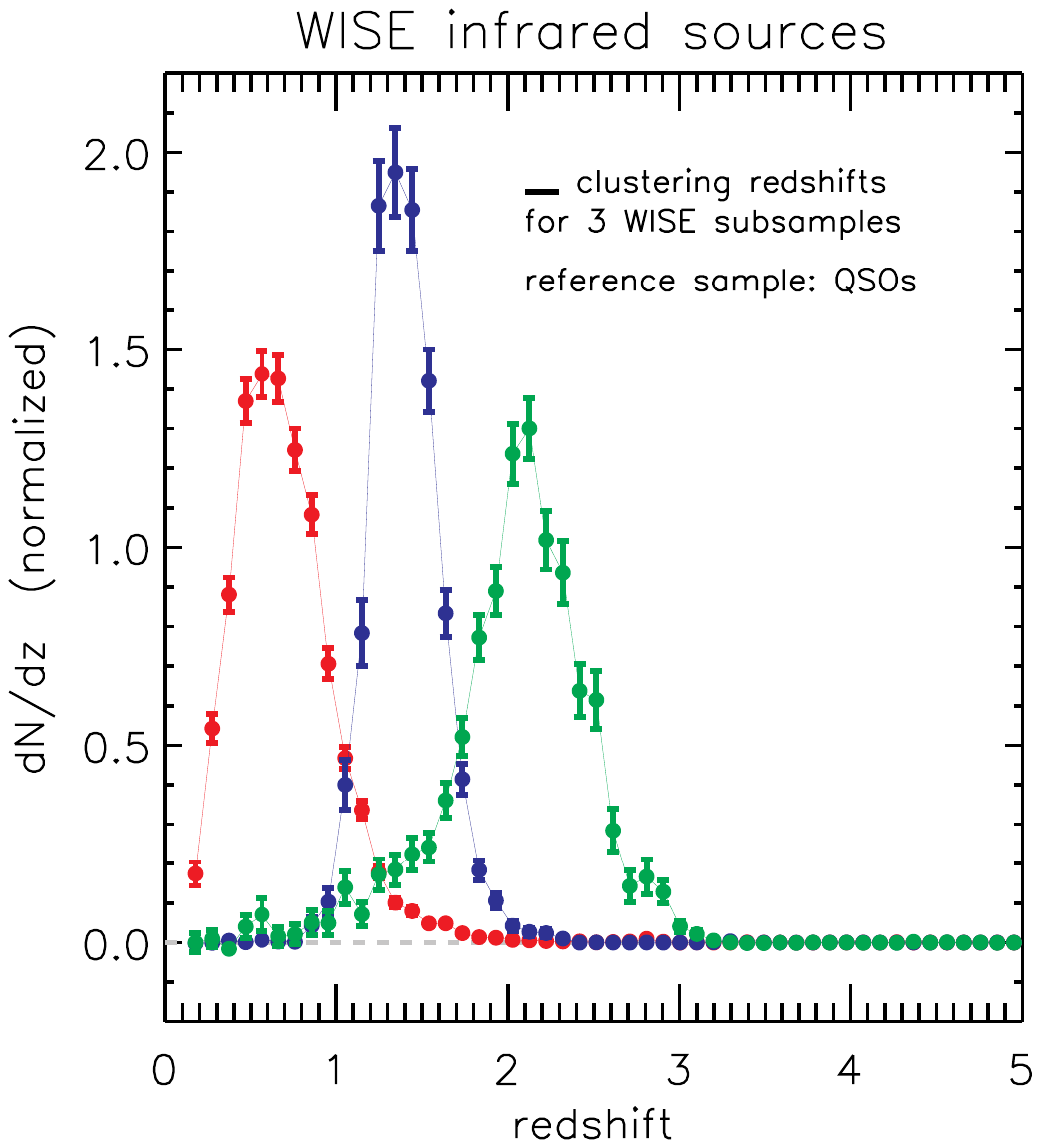}
\includegraphics[width=10cm]{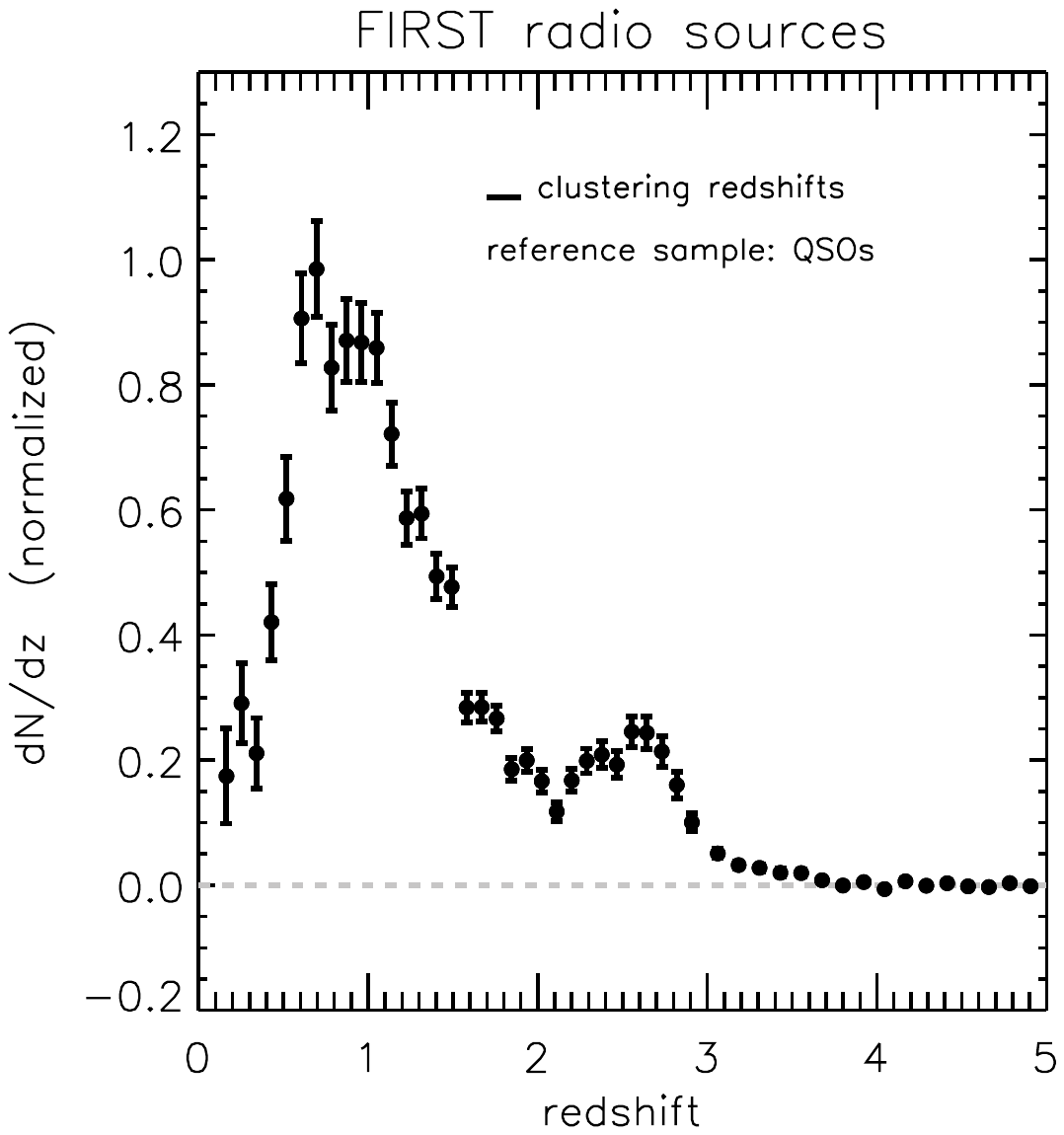}
\caption{
\emph{Left:~} Redshift distributions ${\rm dN/d}z$ (normalized to unity) for three subsamples of WISE sources obtained by measuring their spatial cross-correlation with SDSS quasars. We show the selection criteria for red (Sample 1), blue (Sample 2) and green (Sample 3) samples in Eq.~ \ref{eq:selWISE}.
\emph{Right:~} Redshift distribution of FIRST radio sources obtained by measuring their spatial cross-correlation with SDSS quasars. We observe the existence of sources up to $z\sim3$ as well as a bimodal redshift distribution.
}
%\vspace{-0.5cm}
\label{fig:WISE_FIRST}
\end{figure*}
% . . . . . . . . . . . . . . . . . . . . . . . . . . . . . . . . . .

We now apply our method to estimate the redshift distribution of several populations: (i) Luminous Red Galaxies (LRGs) for which accurate photometric redshifts are available for comparison, (ii) Emission Line Galaxies (ELGs) for which photometric redshift estimation is more difficult to estimate due to the presence of strong emission lines, (iii) infrared sources from WISE survey and (iv) radio sources from the FIRST survey, for which photometric redshifts for the single radio flux density are difficult to define. In the first two cases we will use both spectroscopic quasars and \mgii\ absorbers as reference samples, specifically the SDSS DR7 quasar catalog \citep{schneider10} and the DR7 MgII catalog compiled by \citet{2013ApJ...770..130Z}.  These two samples have different bias evolution profiles, so comparing clustering redshift distribution for the same unknown sample is an interesting test to show whether the technique is insensitive to the reference sample's bias.

We measure spatial cross-correlations between each `unknown' sample and the two spectroscopic populations, integrating over physical scales ranging from zero to 1 Mpc, using a simple weight function $W(\theta)\propto 1/\theta$. 
%Our estimator for the redshift distribution ${\d N/\d} z$ is simply normalized according to Eq.~\ref{eq:normalization}. 
Our goal here is not to construct an optimal estimator but to demonstrate that this technique provides us with a new type of information on redshift distributions, independent of what is obtained through photometric redshifts.

When the inferred redshift distribution is broad, we need to take into account the redshift dependence of the bias of the reference population. For these estimations, we use only our quasar sample, taking our bias evolution from \cite{porciani06}:
\begin{equation}
b_{\rm QSO}(z) = \frac{1}{\sigma_8}\,
\left[
1+\left(\frac{1+z}{2.5} \right)^\gamma
\right]
\label{eq:qso_bias}
\end{equation}
with $\gamma=4$ to provide a better fit to the high-redshift quasar clustering measurements \citep{2012arXiv1212.4526S}.
%
%Throughout the rest of the section we will present normalized redshift distributions $\phi(z)$ according to
%\begin{equation}
%\phi(z)=
%\frac{1}{{\rm N_u}}\,\frac{{\rm d N_u}}{\d z}(z)\;.
%\end{equation}

\subsection{Luminous Red Galaxies}

%The catalogue is selected from the imaging data of the Sloan Digital Sky Survey (SDSS) Data Release 4. 
 
We now apply our technique to the MegaZ-LRG sample \citep{collister07}. This catalogue contains about one million SDSS Luminous Red Galaxies with robust photometric redshifts. This sample spans the  redshift range $0.4 < z < 0.7$ with limiting magnitude $i < 20$. 
The 2dF-SDSS LRG and Quasar \citep[2SLAQ;][]{cannon06} spectroscopic redshift catalogue of 13 000 intermediate-redshift LRGs provides a photometric redshift training set, indicating that the rms photometric redshift accuracy obtained for an evaluation set selected from the 2SLAQ sample is $\sigma_z= 0.049$ averaged over all galaxies. The distribution of photometric redshifts is shown in Figure~\ref{fig:LRG} with the solid line.

We measure the spatial cross-correlation between LRGs and quasars as a function of redshift, and use it to estimate the LRG redshift distribution. The results are shown with the black data points. They demonstrate that the overall shape of the LRG redshift distribution is properly recovered. In addition, the results show that the megaZ-LRG sample is not significantly contaminated by galaxies at other redshifts in the range probed by the quasars.
We then repeat our measurement replacing the quasars with \mgii\ absorbers. The results, as shown in the right panel of Figure~\ref{fig:LRG}, are again in good agreement with the photometric redshift distribution. This provides us with an estimate independent from that obtained with the quasars and shows that different reference samples can be used to obtain consistent results.

\subsection{Emission Line Galaxies}

We now apply our redshift estimation technique to the so-called Emission Line Galaxies (ELGs) from the SDSS \citep{comparat13}.
This corresponds to a sample of faint blue galaxies for which the broad band colors are dominated by emission lines. Following these authors, we have selected the galaxies from the SDSS DR7 database with:
\begin{eqnarray}
\label{eq:ELG}
i&<&21.5 
\\
g-r & < & 1.0
\nonumber\\
r-i & > & -0.917\,(g-r)+0.683
\nonumber\\
r-i & > & 0.5\,(g-r)+0.4\;.
\nonumber
\end{eqnarray}
Using SDSS DR7, this provides us with a sample of about 2.6 million galaxies. We measure the spatial cross-correlation between these sources and quasars as a function of redshift and use it to estimate the redshift distribution of the population. The results are shown in Figure~\ref{fig:ELG} with the black data points. They indicate that the sources selected according to Eq.~\ref{eq:ELG} have a bi-model redshift distribution, with a main population located at $z\sim0.6$ and a second group located at lower redshift.

We also measure the spatial cross-correlation between ELGs and \mgii\ absorbers as a function of redshift. Again, the estimated redshift distribution is in good agreement with that obtained from the spectroscopic quasars. In this case, the overall normalization given by Eq.~\ref{eq:normalization} does not properly apply as the spectroscopic redshift coverage is not wide enough to probe the redshifts of all unknown sources. As a result, the amplitude of ${\rm dN_u/d}z(z\sim0.7)$ obtained with the \mgii\ absorber systems is higher than that the more correct one obtained with quasars as the reference population.\\

Because the redshift distibution is not simple and we are most likely observing two distinction populations of galaxies with different biases, we cannot accurately quantify the relative numbers of the low and high redshift populations. As indicated in section~\ref{}, by applying the clustering redshift technique in subsamples of the population selected in Eq.~\ref{eq:ELG}, one may be able to find a region of the photometric space selecting either the low or high redshift peaks of the distribution. This can be done empirically, without any knowledge of the nature and/or spectral energy distribution of the sources.

\subsection{The WISE infrared survey}

We now apply the clustering redshift technique to a dataset from the Wide-Field Infrared Survey Explorer \citep[WISE;][]{wright10}, a mid-infrared survey satellite which provides us with all-sky observations in four bands, centered at 3.4, 4.6, 12, and 22 $\mu$m (W1 to W4, hereafter). As an illustration we select sources with a magnitude cut $[W_1]<16.5$ and three different selections in color space:
%, $([W_2] - [W_3]) - ([W_1]-[W_2])$:
\begin{eqnarray}
\mathrm{Sample~}1: & \quad 2<[W_{2-3}]<2.5 \nonumber\\
& \quad0.9<[W_{1-2}]<1.2\nonumber\\
\mathrm{Sample~}2: & \quad 2.5<[W_{2-3}]<3 \nonumber\\
& \quad 1.5<[W_{1-2}]<1.8\nonumber\\
\mathrm{Sample~}3: & \quad 3.5<[W_{2-3}]<4 \nonumber\\
& \quad 1.2<[W_{1-2}]<1.5
\label{eq:selWISE}
\end{eqnarray}
where $[W_{i-j}]=[W_i]-[W_j]$. We then cross-correlate these subsamples against the SDSS QSOs as our reference sample. The results are shown in  Figure~\ref{fig:WISE_FIRST}. As can be seen, we observe three distinct redshift distributions, as shown by the different colors: Sample 1 (red), Sample 2 (blue) and Sample 3 (green). These populations appear to have mean redshifts of about 0.5, 1.5 and 2, respectively. While these samples represent only a small fraction of the WISE data, they show that even simple color cuts may be sufficient for selecting non-overlapping samples for cosmological tests.  A future paper will explore the redshift distribution of the WISE data in more detail.

\subsection{The FIRST radio survey}

The Faint Images of the Radio Sky at Twenty centimeters survey \citep[FIRST;][]{becker95} uses the Very Large Array (VLA) to produce a map of the 20 cm (1.4 GHz) sky with a beam size of 5.4\arcsec and an rms sensitivity of about 0.15 mJy/beam. The survey covers an area of about 10,000 deg$^2$ in the north Galactic cap and a smaller area along the celestial equator, both of which roughly coincide with the regions observed by SDSS. With a source surface density of about $90$ deg$^{-2}$, the final catalog includes about one million objects.

Using the SDSS spectrocopic quasar catalog and correcting for bias evolution as given in Equation~\ref{eq:qso_bias}, 
the clustering redshift technique provides us with the redshift distribution shown in Figure~\ref{fig:WISE_FIRST}.  As mentioned in \S\ref{sec:inference}, this distribution has a broad redshift support. We are therefore in a regime substantially departing from our working assumption (Equation~\ref{eq:recovery_criteria}). Hence, without additional knowledge on the redshift evolution of the bias of FIRST objects, we do not expect our redshift distribution estimate to be accurate.  However our results allow us to say with some confidence that the source redshift distribution extends to $z\sim3$ and that there exists two distinct populations of sources, one centered around $z \sim 1$ and a higher redshift cohort around $z \sim 2.5$. Selecting these two populations independently is difficult from radio data only, given the lack of additional parameters available in FIRST, but could potentially be done via cross-matching FIRST sources with additional datasets at other wavelengths.

\section{Conclusions}\label{sec:conclusions}

We have presented a method to infer the redshift distribution of arbitrary datasets, based on spatial cross-correlations with a reference population and we have applied it to various datasets across the electromagnetic spectrum. We have shown that this technique is expected to provide an accurate answer when the unknown population is located within a narrow redshift bin. We have also shown that a large range of departures from this ideal regime can still provide us with redshift estimates accurate enough for numerous applications. For example, at $z<1$, we expect to estimate the mean redshift of color-selected galaxy populations with an uncertainty of $\delta z\sim 0.01$. We have shown that this technique provides better results when first applied to photometric subsamples rather than an entire sample as a whole. Previous works exploring the same avenue \citep[e.g.][]{New:08,2008PhRvD..78d3519H, Matt:10,Schu:10,Matt:12,McQ:13} have focused on large scales where galaxies and dark matter are linearly related to each other. Here we advocate the use of clustering measurements on all available scales and discuss the benefits of using small-scale correlations which tend to be less affected by systematics with real data. In a companion paper \citep{schmidt13} we have used numerical simulations to show the robustness and limitations of this approach. 

To demonstrate the potential of this technique, we have applied the proposed method to estimate the redshift distributions of SDSS luminous red galaxies, emission line galaxies, sources from the WISE infrared survey and the FIRST radio survey. For the first two samples, located at low redshift, we have estimated their redshift distributions using both quasars and absorber systems as the reference population and obtained consistent results. For broad or multi-peaked redshift distributions, as is the case with the ELG and FIRST samples, we cannot obtain a reliable redshift distribution estimate. However, we can robustly estimate the redshift ranges over which the corresponding subsamples exist. More robust redshift distribution estimates can be obtained by applying the clustering redshift technique locally in the space of photometric observables of these datasets. Such a higher level of sophistication will be presented and used  in future analyses. We also note that the clustering redshift technique is a powerful tool to to check for the absence of sources over a given redshift range. This was used in \citep{morrison12} to search for contamination of high redshift Lyman-break galaxies by low redshift interlopers. Finally, we discussed the fact that the ultimate goal of the clustering-redshift technique is to characterize the mapping between the space of photometric observables and redshift space. This characterization can then be used to estimate the clustering-redshift p.d.f. of a single galaxy. 

The application to real data presented in this paper is only a pilot study aimed at demonstrating the potential of the technique which provides us with the ability to characterize the three-dimensional density distribution of sources from the inherently two-dimensional observations of the extragalactic sky. More detailed applications to various surveys will be presented in future papers.\\

\acknowledgments

This work is supported by a NASA grant and the Alfred P. Sloan foundation. RS, SJS and CBM acknowledge the support of NSF Grant AST-1009514. DJ acknowledges the support of DoE SC-0008108 and NASA NNX12AE86G.

Funding for the SDSS and SDSS-II has been provided by the Alfred P. Sloan Foundation, the Participating Institutions, the National Science Foundation, the U.S. Department of Energy, the National Aeronautics and Space Administration, the Japanese Monbukagakusho, the Max Planck Society, and the Higher Education Funding Council for England. The SDSS Web Site is http://www.sdss.org/. 

This publication makes use of data products from the Wide-field Infrared Survey Explorer, which is a joint project of the University of California, Los Angeles, and the Jet Propulsion Laboratory/California Institute of Technology, funded by the National Aeronautics and Space Administration.


\begin{thebibliography}{}

%SDSS DR7
\bibitem[Abazajian et al.(2009)]{abazajain09} Abazajian, K.~N., Adelman-McCarthy, J.~K., Ag{\"u}eros, M.~A., et al.\ 2009, \apjs, 182, 543

\bibitem[Bartelmann \& Schneider (2001)]{bs01} Bartelmann, M., Schneider, P.\ 2001, \physrep, 340, 291

%FIRST Survey
\bibitem[Becker et al.(1995)]{becker95} Becker, R.~H. and White, R.~L. and Helfand, D.~J.\ 1995, \apj, 450, 559

%Benjamin 2013
\bibitem[Benjamin et al.(2010)]{Benj:10} Benjamin J., van Waerbeke L., M\'{e}nard B., Kilbinger M., 2010, \mnras ,408, 1168

%2SLAQ LRGs
\bibitem[Cannon et al.(2006)]{cannon06} Cannon, R., Drinkwater, M., Edge, A., et al.\ 2006, \mnras, 372, 425

\bibitem[Collister et al.(2007)]{collister07} Collister, A., Lahav, O., Blake, C., et al.\ 2007, \mnras, 375, 68 

\bibitem[Comparat et al.(2013)]{comparat13} Comparat, J., Kneib, J.-P., Escoffier, S., et al.\ 2013, \mnras, 428, 1498 

\bibitem[Ho et al.(2008)]{2008PhRvD..78d3519H} Ho, S., Hirata, C., Padmanabhan, N., Seljak, U., \& Bahcall, N.\ 2008, \prd, 78, 043519 


\bibitem[Matthews \& Newman (2010)]{Matt:10} Matthews D. J., Newman J. A., 2010, \apj, 721, 456

\bibitem[Matthews \& Newman (2012)]{Matt:12} Matthews D. J., Newman J. A., 2012, \apj, 745, 180

\bibitem[McQuinn \& White (2013)] {McQ:13} McQuinn, M., \& White, M\ 2013, arXiv:1302.0857

\bibitem[M\'{e}nard (2010)]{menard10} M{\'e}nard, B., Scranton, R., Fukugita, M., Richards,, G.\ 2010, \mnras, 405, 1025

\bibitem[Morrison et al. (2012)]{morrison12} Morrison, C.~B., Scranton, R., M{\'e}nard, B., et al.\ 2012, \mnras, 426, 2489

\bibitem[Narayan (1989)]{narayan89} Narayan, R.\ 1989, \apjl, 339, L53
 
\bibitem[Newman (2008)]{New:08} Newman J. A., 2008, \apj, 684, 88 

\bibitem[Phillipps et al. (1985)]{Phil:85} Phillipps, S.\ 1985, \mnras, 212, 657 

\bibitem[Peebles(1993)]{peebles93} Peebles, P.~J.~E.\ 1993, 
Principles of Physical Cosmology by P.J.E.~Peebles.~Princeton University 
Press, 1993.~ISBN: 978-0-691-01933-8,  

\bibitem[Porciani \& Norberg(2006)]{porciani06} Porciani, C., \& Norberg, P.\ 2006, \mnras, 371, 1824 

\bibitem[Schmidt et al.(2013)]{schmidt13} Schmidt, S., M{\'e}nard, B., Scranton, R., Morrison, C., \& McBride, C.\ 2013, arXiv:1303.0292 

%Schulz 2010
\bibitem[Schulz (2010)]{Schu:10} Schulz A. E., 2010, \apj, 724, 1305

\bibitem[Scoville et al.(2007)]{2007ApJS..172....1S} Scoville, N., Aussel, 
H., Brusa, M., et al.\ 2007, \apjs, 172, 1 


%SDSS QSO catalog
\bibitem[Schneider et al.(2010)]{schneider10} Schneider, D.~P., Richards, G.~T., Hall, P.~B., et al.\ 2010, \aj, 139, 2360

\bibitem[Scranton et al.(2005)]{scranton05} Scranton, R., M{\'e}nard, B., {Richards}, G.~T., et al.\ 2005, \apj, 633, 589

\bibitem[Shen et al.(2012)]{2012arXiv1212.4526S} Shen, Y., McBride, C.~K., White, M., et al.\ 2012, arXiv:1212.4526 

%WISE Survey
\bibitem[Wright et al.(2010)]{wright10} Wright, E.~L., Eisenhardt, P.~R.~M., Mainzer, A.~K., et al.\ 2010, \aj, 140, 1868

\bibitem[York et al.(2000)]{2000AJ....120.1579Y} York, D.~G., Adelman, J., 
Anderson, J.~E., Jr., et al.\ 2000, \aj, 120, 1579 

\bibitem[Zehavi et al.(2011)]{2011ApJ...736...59Z} Zehavi, I., Zheng, Z., Weinberg, D.~H., et al.\ 2011, \apj, 736, 59 


\bibitem[Zhu \& M{\'e}nard(2013)]{2013ApJ...770..130Z} Zhu, G., \& M{\'e}nard, B.\ 2013, \apj, 770, 130 

\end{thebibliography}
\end{document}